\documentclass[prb,twocolumn,showpacs,floats,superscriptaddress]{revtex4}

\usepackage{graphicx}
\usepackage{amsmath}
\usepackage{amssymb}
\usepackage[colorlinks=true,citecolor=blue,linkcolor=blue]{hyperref}
\usepackage{amsfonts}
\usepackage[latin9]{inputenc}
\usepackage{color}
\usepackage{bm}
\usepackage{mathrsfs}
\usepackage{float}
\makeatletter

\newcommand{\Rmnum}[1]{\expandafter\@slowromancap\romannumeral #1@}
\makeatother
\begin{document}

\title{Tunable electronic band structures and zero-energy modes of
heterosubstrate-induced graphene superlattices}
\date{\today }
\author{Xiong Fan}
\affiliation{Department of Physics, Beijing Normal University, Beijing 100875, China}
\author{Wenjun Huang}
\affiliation{Department of Physics, Beijing Normal University, Beijing 100875, China}
\author{Tianxing Ma}
\email{txma@bnu.edu.cn}
\affiliation{Department of Physics, Beijing Normal University, Beijing 100875, China}
\affiliation{Beijing Computational Science Research Center, Beijing 100193, China}
\author{Li-Gang Wang}
\affiliation{Department of Physics, Zhejiang University, Hangzhou 310027, China}

\begin{abstract}
 We propose a tunable electronic band gap and zero-energy modes in periodic
heterosubstrate-induced graphene superlattices. Interestingly, there is an approximate linear relation between the band gap and the proportion of an inhomogeneous substrate (i.e., percentages of different components) in the proposed superlattice, and the effect of structural disorder on the relation is discussed. In an inhomogeneous substrate with equal widths, zero-energy states emerge in the form of Dirac points by using asymmetric potentials, and the positions of Dirac points are addressed analytically. Further, the Dirac point exists at $\mathbf{k}=\mathbf{0}$ only for specific potentials; every time it appears, the group velocity vanishes in the $k_y$ direction, and the resonance occurs. For general cases of an inhomogeneous substrate with unequal widths, part of the zero-energy states are described analytically, and differently,  they are not always Dirac points. Our prediction may be realized on a heterosubstrate such as SiO$_2$/BN.
\end{abstract}


\maketitle
\section{Introduction}\label{part1}
Since graphene was isolated in 2004 \cite{Novoselov2004}, it has attracted great attention \cite{Beenakker2008,Castro2009,Das2011,Goerbig2011,Basov2014,Ma2010,Ni2014}. Graphene is the
true two-dimensional ($2$D) material with one-atom thickness, and the $2$D electronic gas in graphene obeys the massless Dirac equation \cite{Novoselov2005,Zhang2005}. In the past decade, there have been many interesting discoveries in graphene, such as the quantum anomalous Hall effect \cite{Novoselov2005,Purewal2006}, Klein tunneling \cite{Klein1929,Katsnelson2006}, plasmons \cite{Ni2015}, ballistic charge transport \cite{Miao2007,Du2008}, and minimum conductivity \cite{Geim2007}.
Despite these exotic properties, pristine graphene is gapless in the electronic spectrum, which hinders its application in some electronic devices like transistors.

In recent years, many methods have been proposed for opening a gap in the electronic spectrum of graphene \cite{Hicks2013,Magda2014,Baringhaus2014,Palacio2015,Son2006a,Son2006b,Barone2006,Han2007,Li2009,Giovannetti2007,Zhou2007,Geim2007,Kim2008,Balog2010,Song2013, Bokdam2014,Jung2015}. For instance, by breaking A and B sublattice symmetry by using graphite, hexagonal boron nitride (h-BN), and SiC, gaps of $\simeq 10$ meV \cite{Li2009}, $53$ meV \cite{Giovannetti2007}, and $260$ meV \cite{Zhou2007}, respectively, have already been experimentally demonstrated. Moreover, the tunable band gap and electronic properties in the monolayer heterostructure of hexagonal boron nitride and graphene (h-BNC) have been studied intensively \cite{Drost2014,Shinde2011,Zhao2012,Jiang2011,Liu2015}. One of the interesting and heuristic results in h-BNC is that the percentage of h-BN doped in the graphene layer can be used to tune the band gap \cite{Shinde2011}.

Zero-energy modes, critical states in graphene's band structure, always lead to interesting and significant properties. Pristine graphene shows minimum conductivity and the highest shot noise at the unique Dirac point \cite{Katsnelson2006b,Tworz2006}. It is found that the periodic potential induces new Dirac points, and resonance occurs for specific conditions \cite{Brey2009,Park2009,Barbier2010}. It is shown that confinement of the charge carriers in graphene by electrostatic potentials is possible for zero-energy states \cite{Downing2011,Downing2015}.  Most recently, Ferreira and  Mucciolo studied vacancy-induced zero-energy modes in graphene, and they proved that the early field-theoretical picture for the BD\Rmnum{1} class \cite{BDI} is valid well beyond its controlled weak-coupling regime \cite{Ferreira2015}. Further important progress reported recently is that the presence of Majorana zero-energy modes is predicted in graphene/superconductor junctions \cite{San-Jose2015}.

Inspired by these studies, our motivation here is to study the electronic band structure of heterosubstrate-induced graphene superlattices (GSLs), which refer to the GSLs with periodically modulating substrates via the BN (or SiC, graphite) and SiO$_{2}$ substrates. It is known that the behavior of electrons in graphene may be different on different substrates; thus it is expected that the electronic properties in such heterosubstrate-induced GSLs can be tunable. It would be useful to mention that such heterosubstrate-induced GSLs could be fabricated technically by many advanced growth methods in electronics, like those in Refs. \onlinecite{Tsu2005,Meyer2008,Marchini2007,Coraux2008}.

In this paper, we present a theory of tunable electronic band structures and zero-energy modes in the heterosubstrate-induced GSLs. It is interesting to see that the electronic band gap of such heterosubstrate-induced GSLs is dependent on the width ratio of different substrates. More importantly, this paper further shows further shows that there exist zero-energy states in such heterosubstrate-induced GSLs, which may be controlled by applying unsymmetrical square potentials. These results are compared with those in the gapless GSLs\cite{Brey2009,Park2009,Barbier2010,Wang2010,Ma2012} and the gapped GSLs \cite{Wang2011}. The properties of the electronic transport such as the conductivity and the Fano factor are discussed in detail. The possible realization of our results in experiment is also suggested.

The outline of this paper is as follows. In Sec.\,\ref{part2}, we introduce our model and method for calculating band structures, electronic transmission, conductivity, and the Fano factor. In Sec.\,\ref{part3}, we show the properties with different parameters and discuss the effect of lattice constants and structural disorder on the band gap. Afterward, we investigate the appearance of zero-energy states, the group velocities, and the resonances near zero energy. Finally, in Sec.\,\ref{part4}, we summarize our results and draw our conclusions.

\section{Model and formula}\label{part2}
\begin{figure}[t b]
\centering
\includegraphics[width= 7cm]{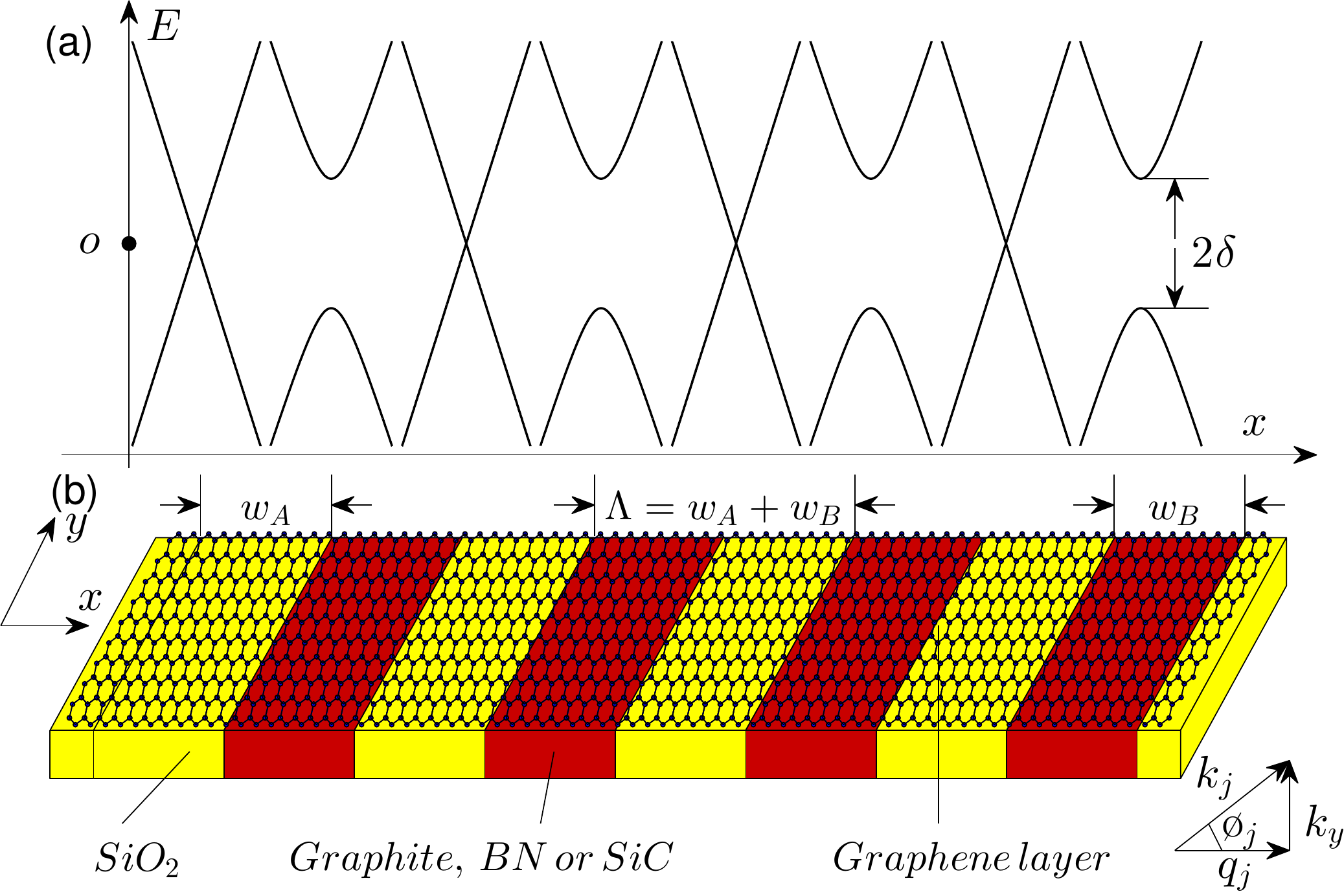}
\caption{(Color online) (a) Schematic diagram of the electronic spectrums of
gapless and gapped graphene without potentials. (b) Schematic
of the monolayer graphene superlattice on the periodic heterosubstrate. $2\delta$ is the band gap induced by the graphite, BN, or SiC substrate; $w_{A(B)}$ is the width of the SiO$_2$ (graphite, BN, and SiC) substrate, as well as the width of gapless (gapped) graphene; and $\Lambda=w_A+w_B$ is the lattice constant. $k_j$ is the wave vector in the $j$th region, $q_j$ and $k_y$ are components of $k_j$ in the $x$ and $y$ directions, and $\o_j$ is the transporting angle in the $j$th region.}
\label{fig1}
\end{figure}

In this model, gapless and gapped graphene are periodically hybridized, such as that shown in Fig. \ref{fig1}. The gapless graphene is fabricated on the SiO$_{2}$ substrates, and the gapped graphene is grown on the h-BN substrates \cite{Kindermann2012,Giovannetti2007}. The difference between gapless and gapped graphene is whether an electronic band gap caused by the sublattice symmetry breaking exists. In gapless graphene, there is a famous linear band structure with a unique Dirac point, whereas, in gapped graphene, a $2\delta$-wide band gap near zero energy exists.

We assume the length of these cells along the $y$ direction is infinite, and we call such a periodic superlattice, shown in Fig. \ref{fig1}, a heterosubstrate-induced GSL. $w_A$ is the width of the gapped-graphene subcell on the SiO$_2$ substrates, $w_B$ is the width of the gapped-graphene subcells on BN/SiC substrates, and $\Lambda=w_A+w_B$ is the lattice constant of the whole periodic structure.

\begin{figure*}[t b]
\includegraphics[width= 13cm]{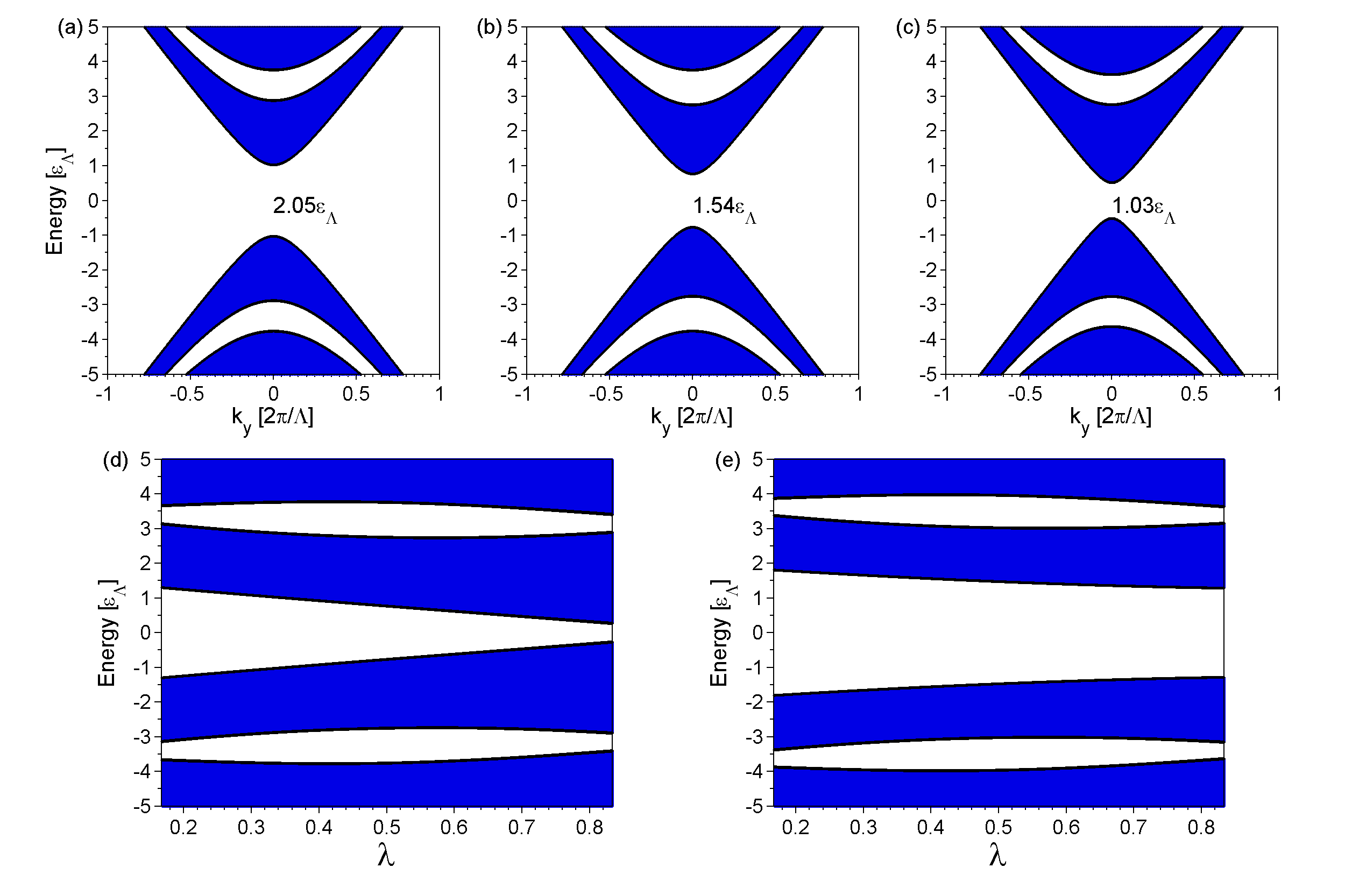}
\caption{(Color online) (a)-(c) Electronic band structures for (a) $\protect\lambda=0.33$, (b) $\protect\lambda=0.50$, and (c) $\protect\lambda=0.67$. (d) and (e) The dependence of band structures on $\lambda$ under (d) $k_y=0$ and (e) $k_y=0.2$ (in units of $2\protect\pi/\Lambda$). In all cases, $V_A=V_B=0$ and $\protect\delta=\protect\pi\protect\varepsilon_\Lambda/2$. Here energy is in units of $\protect\varepsilon_\Lambda \equiv \hbar \protect \upsilon_F/\Lambda$, $\protect\lambda= w_A/\Lambda$, and $\Lambda=60 $ nm.}
\centering
\label{fig2}
\end{figure*}

For both gapless and gapped monolayer graphene, the charge carrier near the $K$ point can be universally described by the Hamiltonian
\begin{equation}  \label{H}
\hat{H}=\upsilon _{F}\bm{\sigma}\cdot {\mathbf{p}}+V(x)\hat{I}+\delta\sigma_z,
\end{equation}
where $\upsilon _{F}\simeq10^{6}$ m/s is the Fermi velocity, ${\bm{\sigma}} =\left({\sigma}_{x},{\sigma}_{y}\right)$, and $\sigma_x$, $\sigma_y$, and $\sigma_z$ are Pauli matrices; $\mathbf{p}=(p_x, p_y)=(-i\hbar\frac{\partial}{\partial_x},-i\hbar\frac{\partial}{\partial_y})$ is the two-component momentum operator, $V(x)$ is the one-dimensional ($1$D) square potential depending on the $x$ direction, $\hat I$ is a $2\times 2$ unit matrix, and $\delta$ is the half width of the band gap opened by the sublattice symmetry breaking. When $\delta=0$, the Hamiltonian (1) describes gapless graphene. The Hamiltonian $\hat H$ acts on the two-component pseudospin wave function $\Psi=(\tilde{\psi}_A,\tilde{\psi}_B)$, where $\tilde{\psi}_A$ and $\tilde{\psi}_B$ are the smooth envelope functions for two triangular sublattices in monolayer graphene and $\tilde{\psi}_{A,B}$ can be written as $\psi_{A,B}(x)e^{ik_yy}$ due to the translation invariance. The solution of the eigen-equation $\hat{H}$ leads to the transfer matrix \cite{Wang2010,Ma2012}
\begin{equation}
M_{j}(\Delta x,E,k_{y})=\left(
\begin{array}{cc}
\frac{\cos (q_{j}\Delta x-{\o} _{j})}{\cos {\o} _{j}} & i\frac{\sin
(q_{j}\Delta x)}{p_{j}\cos {\o} _{j}} \\
i\frac{p_{j}\sin (q_{j}\Delta x)}{\cos {\o} _{j}} & \frac{\cos (q_{j}\Delta
x+{\o} _{j})}{\cos {\o} _{j}}
\end{array}
\right) \ ,  \label{tranm1}
\end{equation}
which connects the wave functions at $x$ and $x+\Delta x$ inside the $j$th potential. Here in Eq. (\ref{tranm1}),\\
$q_j=\left\{\begin{matrix}
\quad\text{sgn}(\eta_{j+})\sqrt{{k_{j}}^2-{k_y}^2},\quad{k_{j}}^2>{k_{y}}^2,\\
i\sqrt{{k_y}^2-{k_j}^2}, \quad\text{otherwise},
\end{matrix}\right.$\\
is the $x$ component of the wave vector inside the $j$th potential,
$\sin\,{\o_j}=k_y/k_j$,
$\cos\,{\o_j}=q_j/k_j$,
$p_j=\eta_{j-}/k_j$,
and the wave vector inside the potential $V_j$ can be expressed as\\
$k_j=\left\{\begin{matrix}
\quad\text{sgn}(\eta_{j+})[(E-V_j)^2-\delta^2]/(\hbar\upsilon_F),\quad|E-V_j|>\delta,\\
i[\delta^2-(E-V_j)^2]^{1/2}/(\hbar\upsilon_F),\quad\text{otherwise},
\end{matrix}\right.$,\\
where $\eta_{\pm}\equiv[E-V(x)\pm\delta]/(\hbar \upsilon_F)$.
${\o_j}$ is regarded as the transporting angle in the $j$th region, and it should be noted that the angle ${\o_j}$ is not always a
real number because the evanescent mode exists. In the case of $\eta_{j+}=0$, Eq. (\ref{tranm1}) is replaced by \cite{Wang2011}
\begin{equation}
M_{j}(\Delta x,E,k_{y})=\left(
\begin{array}{cc}
\exp(k_y\Delta x) & 0 \\
ip_j\sinh(k_y\Delta x) & \exp(-k_y\Delta x)
\end{array}
\right) \ ,  \label{tranm2}
\end{equation}
where $p_j=\eta_{j-}/k_y$. When $\eta_{j-}=0$, Eq. (\ref{tranm1}) is replaced by \cite{Wang2011}
\begin{equation}
M_{j}(\Delta x,E,k_{y})=\left(
\begin{array}{cc}
\exp(k_y\Delta x) & ip_j\sinh(k_y\Delta x) \\
0& \exp(-k_y\Delta x)
\end{array}
\right) \ ,  \label{tranm3}
\end{equation}
where $p_j=\eta_{j+}/k_y$. The above results are also valid for gapless graphene when $\delta=0$.

For an infinite periodic system $(AB)^N$, with $N\rightarrow\infty$, the symbols $A$ and $B$ denote gapless and gapped graphene with square potentials $V_A$ and $V_B$, respectively.  The wave function of this periodic system is the Bloch wave function. Therefore the electronic dispersion relation is governed by
\begin{equation}  \label{bloch1}
\cos (\beta_x\Lambda)=\frac{1}{2}\quad\text{Tr}[M_AM_B],
\end{equation}
where $\beta_x$ is the $x$ component of the Bloch wave vector in the whole system and $\Lambda$ is the lattice constant labeled in Fig. \ref{fig1}(b). If $\beta_x$ has a real solution, there is an electron (hole) state in the band structure; otherwise, there is a band gap. For general cases of $\eta_{j+}\eta_{j-}\neq 0$ (i.e., $k_j \neq 0$), substituting Eq. (\ref{tranm1}) into Eq. (\ref{bloch1}), we have
\begin{eqnarray}\label{bloch2}
\cos (\beta_x\Lambda)&=&\cos (q_Aw_A+q_Bw_B)-\sin (q_Aw_A)\sin (q_Bw_B)\notag \\
&&\times\frac{\frac{{p_{B}}^2+1}{p_{B}}-2\cos (\o_A-\o_B)}{2\cos (\o_A)\cos (\o_B)}.
\end{eqnarray}
Equation (\ref{bloch2}) will be used to find zero-energy modes and band structures in the next section. Here $p_A=1$ has been used.

Next, we discuss the wave function, the transmission probability, the conductivity, and the Fano factor (ratio of the shot-noise power and the current) for a finite periodic-potential system, which are a reflection of the band structure for infinite periodic systems. From the continuity of both wave functions $\psi_A$ and $\psi_B$, the electronic transmission and reflection amplitudes can be obtained by \cite{Wang2011}
\begin{eqnarray}  \label{tm1}
t(E,k_{y})=&\frac{2p_{0}\cos \o _{0}}{x_{22}p_{0}e^{-i\o
_{0}}+x_{11}p_{e}e^{i\o _{e}}-x_{12}p_{0}p_{e}e^{i(\o _{e}-\o _{0})}-x_{21}},\notag \\
r(E,k_{y})=&\frac{x_{22}p_{0}e^{-i\o
_{0}}-x_{11}p_{e}e^{i\o _{e}}-x_{12}p_{0}p_{e}e^{i(\o _{e}+\o _{0})}+x_{21}}{x_{22}p_{0}e^{-i\o
_{0}}+x_{11}p_{e}e^{i\o _{e}}-x_{12}p_{0}p_{e}e^{i(\o _{e}-\o _{0})}-x_{21}},
\end{eqnarray}
where $\o_{0(e)}$ is the incident (exit) angle and $x_{ij} (i,j=1,2)$ is the element of $\mathbf{X}=\prod\limits_{j=N}^{1}M_{j}(w_{j},E,k_{y})$, which is the entire transfer matrix from the incident to exit edge. We assume $V_{0(e)}\gg V_{j}$; thus $\o_{0(e)}\simeq 0$. Two components of the electronic wave function are expressed by \cite{Wang2010}
\begin{eqnarray}  \label{wavefunction}
\psi_A(x)&=&\psi_i(E,k_y)[(1+r)Q_{11}+p_{0}(e^{i\o _{0}}-re^{-\o _{0}}Q_{12})],\notag \\
\psi_B(x)&=&\psi_i(E,k_y)[(1+r)Q_{21}+p_{0}(e^{i\o _{0}}-re^{-\o _{0}})Q_{22}],\notag \\
\end{eqnarray}
where $\psi_i(E,k_y)$ is the incident wave packet of the electron at $x=0$ and $Q_{ij}(i,j=1,2)$ is the element of the matrix $Q=M_j(x-x_{j-1},E,k_y)\prod\limits_{i=j-1}^{1}M_i(w_i,E,k_y)$.

With the transmission probability $T=|t|^2$, the conductance and the Fano factor for a given energy can be obtained by \cite{Buttiker1993,Masir2010}
\begin{eqnarray}  \label{FG}
G&=&\frac{4e^2L_y}{{2\pi} h}\int_{-\infty}^{\infty} T(E,k_y)\,dk_y,\notag \\
F&=&\frac{\int_{-\infty}^{\infty}T(E,k_y)[1-T(E,k_y)]\,dk_y}{\int_{-\infty}^{\infty} T(E,k_y)\,dk_y},
\end{eqnarray}
where all degeneracies are included. The conductivity is $\sigma= G\times L_x/L_y$, where $L_{x(y)}$ is the length of the graphene stripe in the $x(y)$ direction.

\section{Result and discussion}\label{part3}
Now, let us use the above equations to calculate the electronic band structures and the properties of transport for different situations.  We would like to point out that the edge effect between gapless and gapped graphene can be neglected when the widths $w_A$ and $w_B$ are sufficiently larger than the sublattice size of graphene.
\subsection{Electronic band structures}\label{part3A}

From Eq. (\ref{bloch1}), we can obtain the electronic band structures for the infinite periodic systems. Now, we let $w_A=\lambda \Lambda$, then $w_B=(1-\lambda)\Lambda$, and $\lambda$ is the proportion of gapless graphene over the whole structure. For convenience, the energy is in units of $\varepsilon_\Lambda\equiv\hbar\upsilon_F/\Lambda$, and the transverse wave vector $k_y$ is in units of $2\pi/\Lambda$.
\begin{figure}[ h b]
\centering
\includegraphics[width=8.7cm]{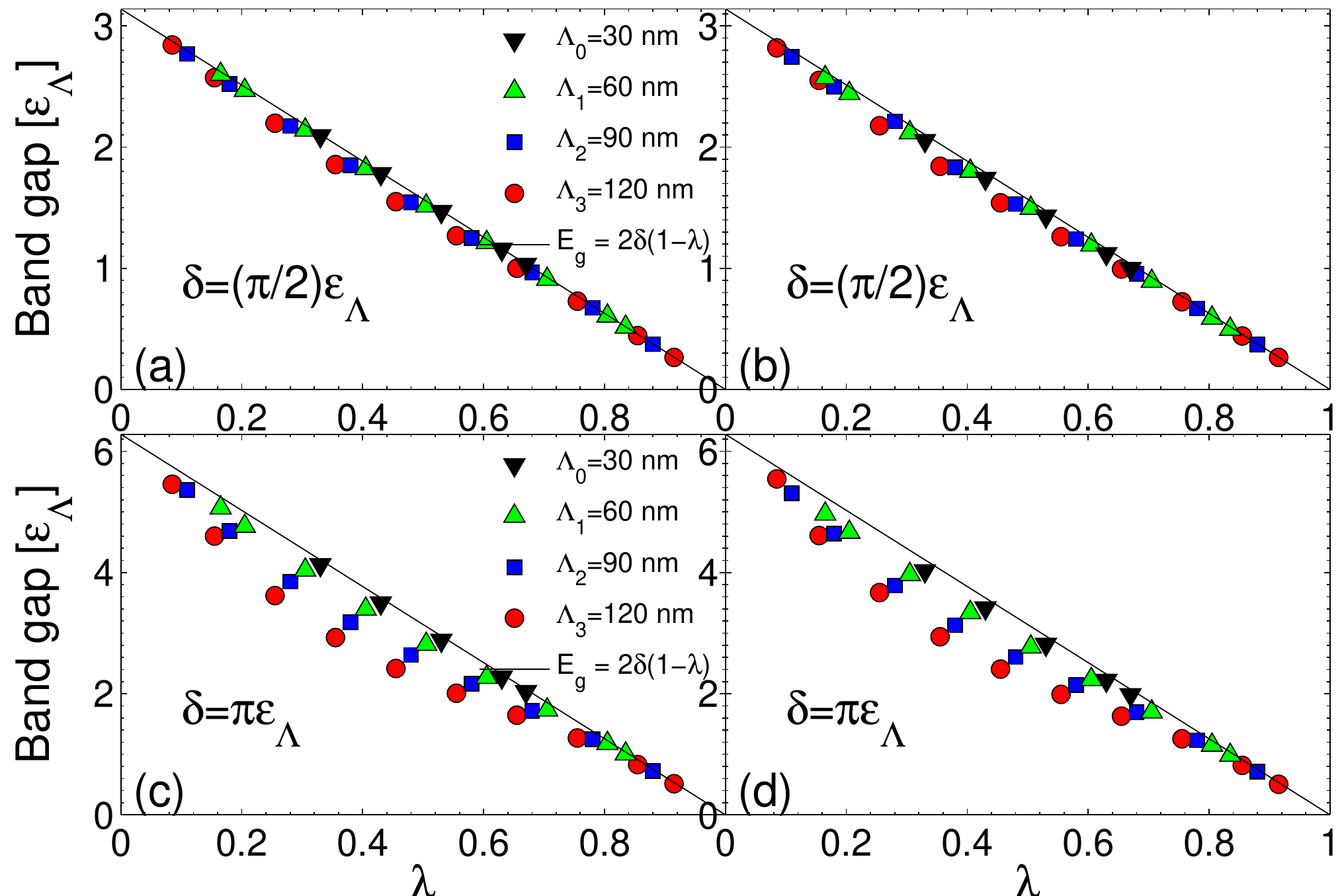}
\caption{(Color online)  (a) and (c) The band gap versus the proportion $\protect\lambda$ with $\protect\lambda=w_A/\Lambda$. (b) and (d) Effect of the structural disorder on the band gap with $w_A=\lambda\Lambda+R_1$ and $w_B=(1-\lambda)\Lambda+R_1$, where $R_1$ is a random number between $-2.5$ and $+2.5$ nm. Here the energy is in units of $\protect
\varepsilon_\Lambda=\hbar\protect\upsilon_F/\Lambda_1$.}
\label{fig3}
\end{figure}

\begin{figure}[!h]
\centering
\includegraphics[width= 8cm]{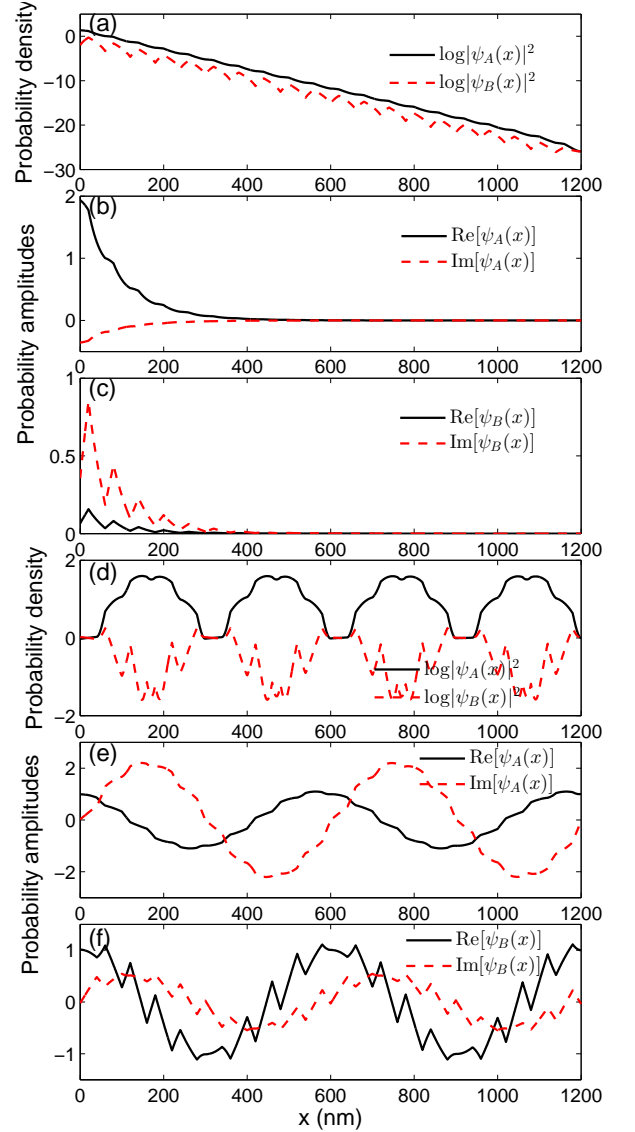}
\caption{(Color online) Evolutions of (a) the probability densities $|\psi_A|^2$ and $|\psi_B|^2$ and (b) and (c) the probability amplitudes $\psi_A$ and $\psi_B$ with $\lambda=0.33$. (d)-(f) Evolutions of probability densities and amplitudes with $\lambda=0.67$. For all cases, $E=0.8 \varepsilon_\Lambda$ and $k_y=0$. The other parameters are the same as those in Fig. \ref{fig2}.}\label{fig4}
\end{figure}
Figure \ref{fig2} shows the band gap around the Fermi level for different $\lambda$. In Figs. \ref{fig2}(a)-\ref{fig2}(c), with a fixed lattice constant, the width of the band gap near the Fermi level changes from $2.05\varepsilon_\Lambda$ to $1.54\varepsilon_\Lambda$ to $1.03\varepsilon_\Lambda$ with $\lambda$ from $0.33$ to $0.50$ to $0.67$. Figures \ref{fig2}(d) and \ref{fig2}(e), with $k_y=0$ and $k_y=0.2$ (in units of $2\pi/\Lambda$), respectively, show that the top of the valence band (the
bottom of the conduction band) increases (decreases) almost linearly as $\lambda$ increases, and the band gap for $k_y=0.2$ is larger than that of the normal case ($k_y=0$). The gap around the Fermi level completely opens compared with the symmetric forbidden bands in the valence band and conduction band [see Figs. \ref{fig2}(a)-\ref{fig2}(c)], and the center position of the gap is robust against both the proportion $\lambda$ [see Figs. \ref{fig2}(a)-\ref{fig2}(e)] and the lattice constant $\Lambda$ (not shown). To find the relation between the width of the band gap and $\lambda$, we plot the band gap's width $E_g$ versus $\lambda$ in Fig. \ref{fig3}. In Figs. \ref{fig3}(a) and \ref{fig3}(c), it is shown that the relations between $E_g$ and $\lambda$ can be approximately described by the linear function $E_g=2\delta(1-\lambda)$. Additionally, both larger $\delta$ and larger $\Lambda$ widen the deviation. In Figs. \ref{fig3}(b) and \ref{fig3}(d), we plot the band gap versus $\lambda$ under the structural disorder. We consider a finite periodic structure $(AB)^{20}$ with a width deviation of $\pm2.5$ nm. From Figs. \ref{fig3}(b) and \ref{fig3}(d), we can find that the relations are robust against the structural disorder. Early works focus on the band gap of graphene nanoribbons (GNRs). In contrast, the band gap of a GNR scales inversely with the channel width \cite{Son2006a,Barone2006}, and it has been proved by experiment \cite{Han2007}. Results similar to those in GNRs are also observed in silicene nanoribbons \cite{Song2010}.
\begin{figure}[t]
\includegraphics[width= 8.5cm]{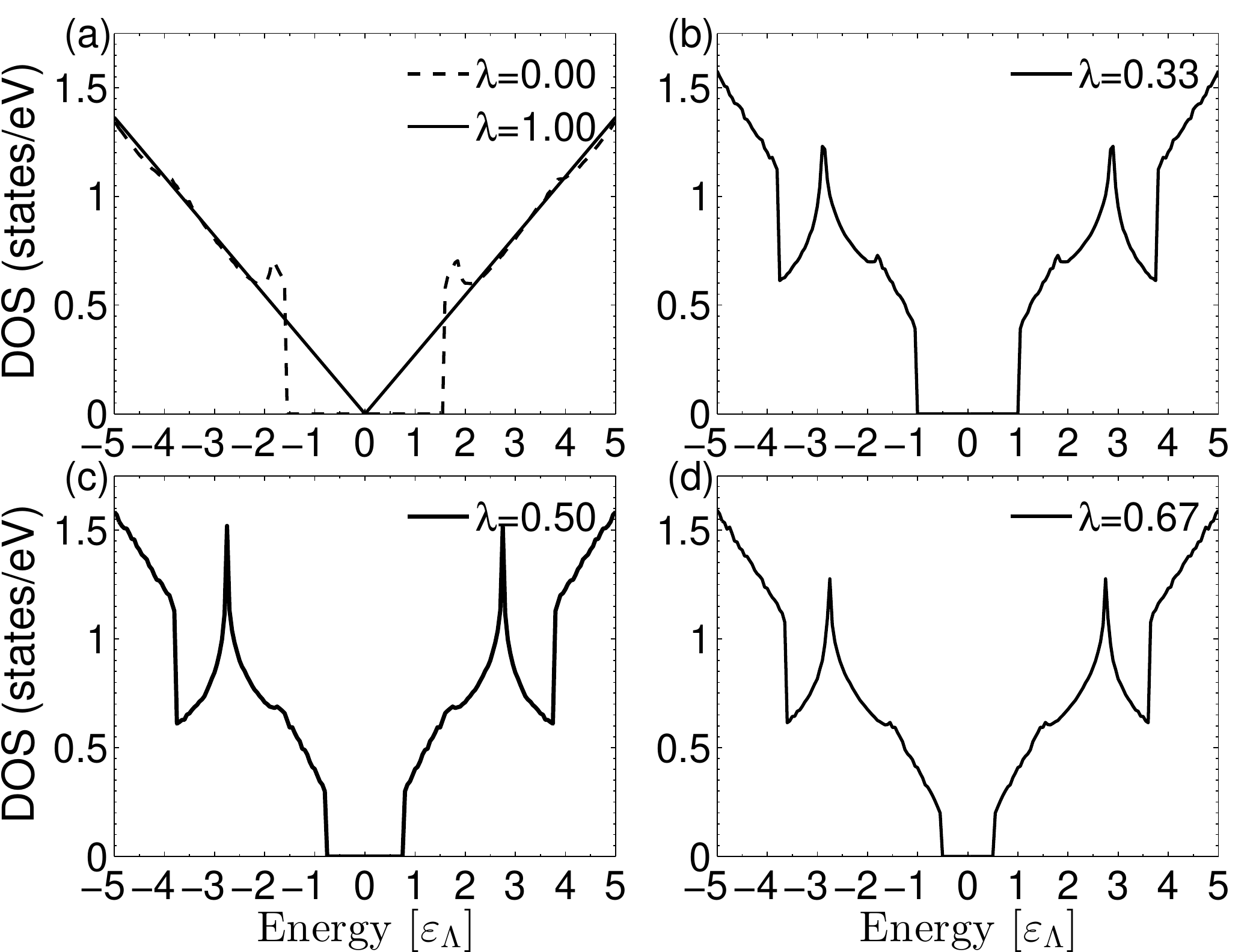}
\caption{(Color online) Density of states vs energy for different $\lambda$. The other parameters are the same as those in Fig. \ref{fig2}.}\label{fig5}
\centering
\end{figure}

Now, we try to qualitatively describe the scaling law of band gaps by introducing competition between propagating and decaying waves in the whole system. It is obvious that electronic energy $E$, interaction potential $\delta$, and proportions of inhomogeneous substrate all affect the competition. Higher $|E|$ ($\delta$) impairs (strengthens) the effect of the decaying wave and strengthens (impairs) the effect of the propagating wave; a larger proportion of gapped graphene also strengthens the effect of the decaying wave and, accordingly, impairs the effect of the propagating wave. In Fig. \ref{fig3}, with fixed $\delta=(\pi/2)\varepsilon_{\Lambda}$, the scaling law of the band gap versus $\lambda$ is almost linear, which may indicate that the impacts of $|E|$ and $\delta$ on the competition are almost equal. When $\delta=\pi\varepsilon_{\Lambda}$, the impact of higher $|E|$ on the competition becomes stronger than the impact of $\delta$; hence the band gap is located under the linear scaling law.
Specifically, we illustrate the evolutions of a decaying-dominated wave and a propagating-dominated wave in Figs. \ref{fig4}(a)-(c) and \ref{fig4}(d)-(f), respectively. With fixed $E=0.8\varepsilon_{\Lambda}$, $k_y=0$, and $\lambda=0.33$ [see Figs. \ref{fig4}(a)-\ref{fig4}(c)], both the transmission probability and amplitude decay quickly as $x$ increases, which results in no electronic state in the band structure, as shown in Fig. \ref{fig2}(a). In contrast, changing $\lambda$ to $0.67$ [see Fig. \ref{fig4}(d)-\ref{fig4}(f)] results in a propagating wave called the Bloch wave in the whole system. Using Eq. (\ref{bloch2}), we find that the Bloch wave vector in the $x$ direction is about $0.01045$ nm$^{-1}$, and the corresponding $x$-component wavelength is $601.3$ nm, which is surely that pictured in Figs. \ref{fig4}(d)-\ref{fig4}(f); thus there is an electronic state in the band structure for $\lambda=0.67$ in Fig. \ref{fig2}(c).

In addition, we can quantitatively verify the scaling law of the band gap by the density of states (DOS), which is plotted in Fig. \ref{fig5}. The DOS for the superlattice system is given by
\begin{equation}  \label{dos}
D(E)=\frac{4S}{{(2\pi)}^2}\sum_n\int_{-\pi/\Lambda}^{\pi/\Lambda}d\beta_x\int_{-\infty}^{\infty}dk_y\delta[E-E_n(\beta_x,k_y)],
\end{equation}
where $S$ is the unit-cell area in the superlattice system. In numerical methods, we substitute a Gaussian for the $\delta$ function to compensate for the discrete $\beta_x$ and $k_y$. Figures \ref{fig5}(a)-\ref{fig5}(d) explicitly show that the DOS is zero in the gap determined by the scaling law for different $\lambda$.
\begin{figure}[t]
\centering
\includegraphics[width=7.5cm]{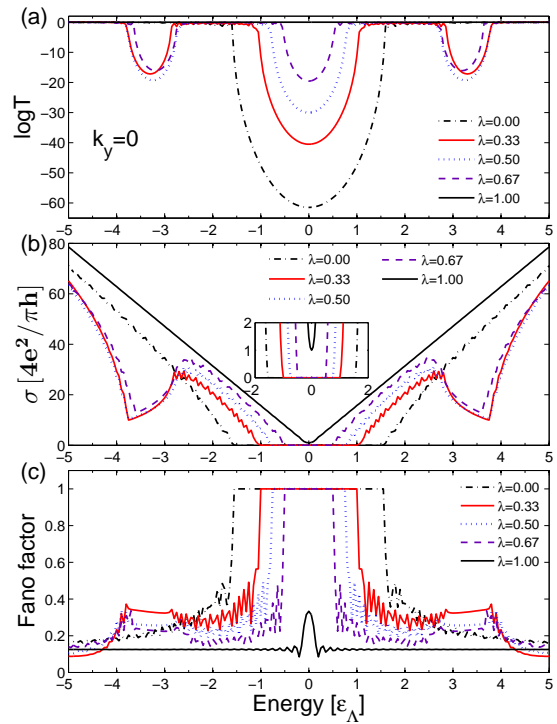}
\caption{(Color online) (a) Transmission probabilities vs the energy of the charge carrier with fixed $k_y=0$.(b) Conductivity and (c) Fano factor vs the energy of the charge carrier. The cases of $\protect\lambda=0.33$, $0.50$, and $0.67$, with the lattice constant $\Lambda=60$ nm and the periodic number $N=20$, correspond to the cases in Figs. \protect\ref{fig2}(a)-\protect\ref{fig2}(c), respectively. For all cases, $V_A=V_B=0$, $L_x=1200$ nm, and $L_y/L_x\gg1$.}
\label{fig6}
\end{figure}

Through Eq. (\ref{FG}) and the equation $\sigma=G\times L_x/L_y$, we can obtain the conductivity and the Fano factor for the finite periodic system. Figure \ref{fig6} shows the transmission probability, the conductivity, and the
Fano factor versus the electronic energy with different $\lambda$. The cases with $\lambda=0.33$, $0.50$, and $0.67$ correspond to the cases in Figs. \ref{fig2}(a)-\ref{fig2}(c), respectively, and $\lambda=0$ and $1$ separately correspond to uniform gapped graphene and uniform gapless graphene. It is seen that in the band gap, the conductivity tends to be zero, and the corresponding Fano factor tends to be the integer $1$. For pure gapless graphene, i.e., $\lambda=1$, the conductivity exhibits the minimum $4e^2/\pi h$ at the Dirac point, and the corresponding Fano factor takes the maximum $1/3$ \cite{Katsnelson2006b,Tworz2006}.
We also notice that outside the band gap, as the absolute value of energy $|E|$ increases, the conductivity will increase almost linearly with small oscillations due to the propagating modes in the finite structures. The additional forbidden band leads to the reduction of the conductivity, which can be checked around the energy $\pm4\varepsilon_\Lambda$ in Figs. \ref{fig2}(a)-\ref{fig2}(c) and \ref{fig6}(b). The corresponding Fano factor gradually changes from 1 to a small value (about $0.1$). This indicates that the transport becomes ballistic.

The above discussions are under the condition of $V_A=V_B=0$; nevertheless, the gate voltage $V_A=V_B=V_{gate}$ can change the band gap originally around the Fermi level $E_F=0$ to the position of $E_F=V_{gate}$. For unequal $V_A$ and $V_B$, a stable gap in the band structure will also exist \cite{zk}. It will be discussed further in the next section.

\subsection{Zero-energy modes}\label{part3B}
\begin{figure}[!htbp]
\centering
\includegraphics[width=7cm]{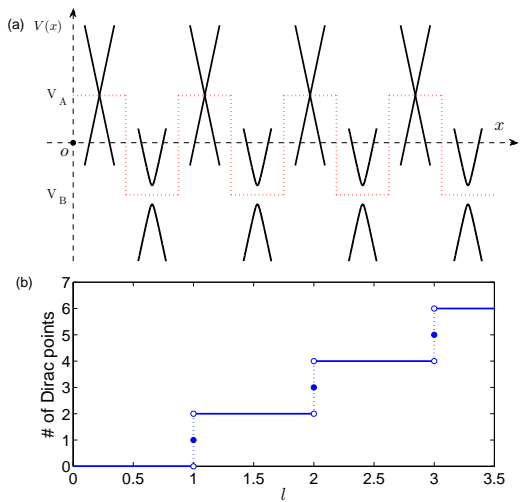}
\caption{(Color online) (a) Schematic diagram of the periodic square potentials (red dotted line) and the electronic spectra of gapless graphene (denoted by large X symbols) and gapped graphene (large V symbols). (b) Number of Dirac
points (spin and valley degeneracies are not included) vs $l$ with $\protect\delta=\protect\pi \protect\varepsilon_\Lambda/2$, $\protect\lambda=0.5$, $V_{A}=2l\protect\pi\protect\varepsilon_\Lambda$, and $V_B$ satisfying Eq. (\protect\ref{VB}). From $V_{A}=2l\protect\pi\protect\varepsilon_\Lambda$, we know that $l$ scales linearly with the potential $V_A$.}
\label{fig7}
\end{figure}

\begin{figure*}[b t p]
\centering
\includegraphics[width=17cm]{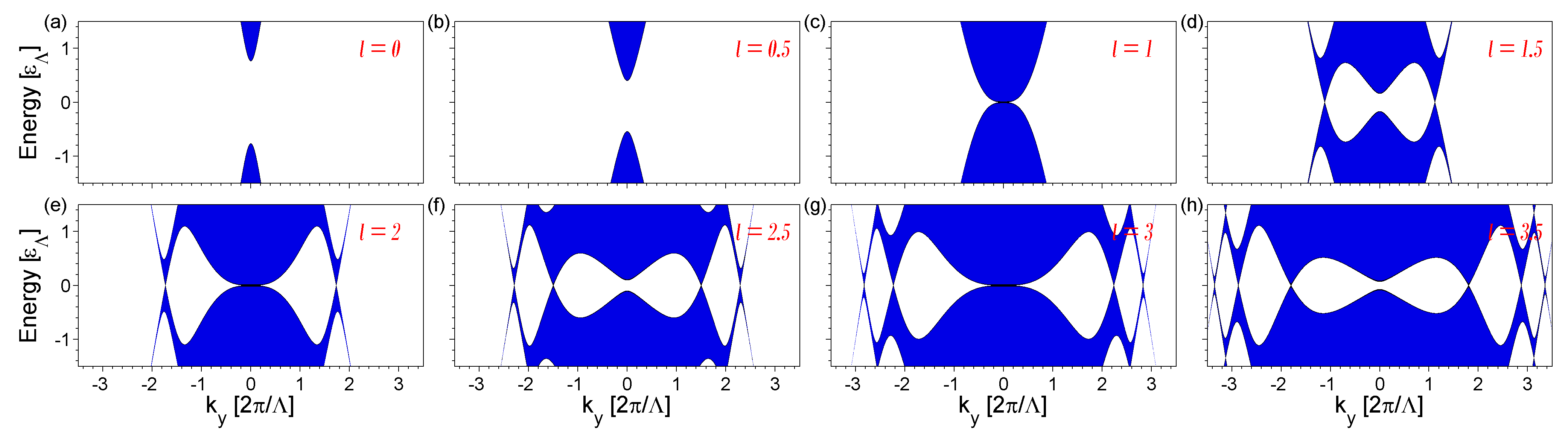}
\caption{(Color online) Electronic band structures for different values of $l$ with $\protect\delta=\protect\pi \protect\varepsilon_\Lambda/2$, $\protect\lambda=0.5$, $V_{A}=2l\protect\pi\protect\varepsilon_\Lambda$, and $V_B$ satisfying Eq. (\protect\ref{VB}). $\Lambda=40$ nm is set in this calculation.}
\label{fig8}
\end{figure*}

In this part, we start with Eq. (\ref{bloch2}) to find zero modes. We discuss the simple and typical case of equal widths of inhomogeneous substrate in detail, i.e., $\lambda=0.5$, so that $w_A=w_B$. Assuming $V_A>V_B$, in the energy range $V_B+\delta<E<V_A$, we have $q_A<0$ and $q_B>0$. When $q_Aw_A+q_Bw_B=0$ in particular, Eq. (\ref{bloch2}) becomes
\begin{equation}  \label{bloch3}
\cos(\beta_x\Lambda)=1+\sin^2(q_Aw_A)\frac{\frac{{p_{B}}^2+1}{p_{B}}-2\cos(\o
_A-\o_B)}{2\cos (\o_A)\cos (\o_B)}.
\end{equation}

\begin{figure}[b t p]
\centering
\includegraphics[width=7.5cm]{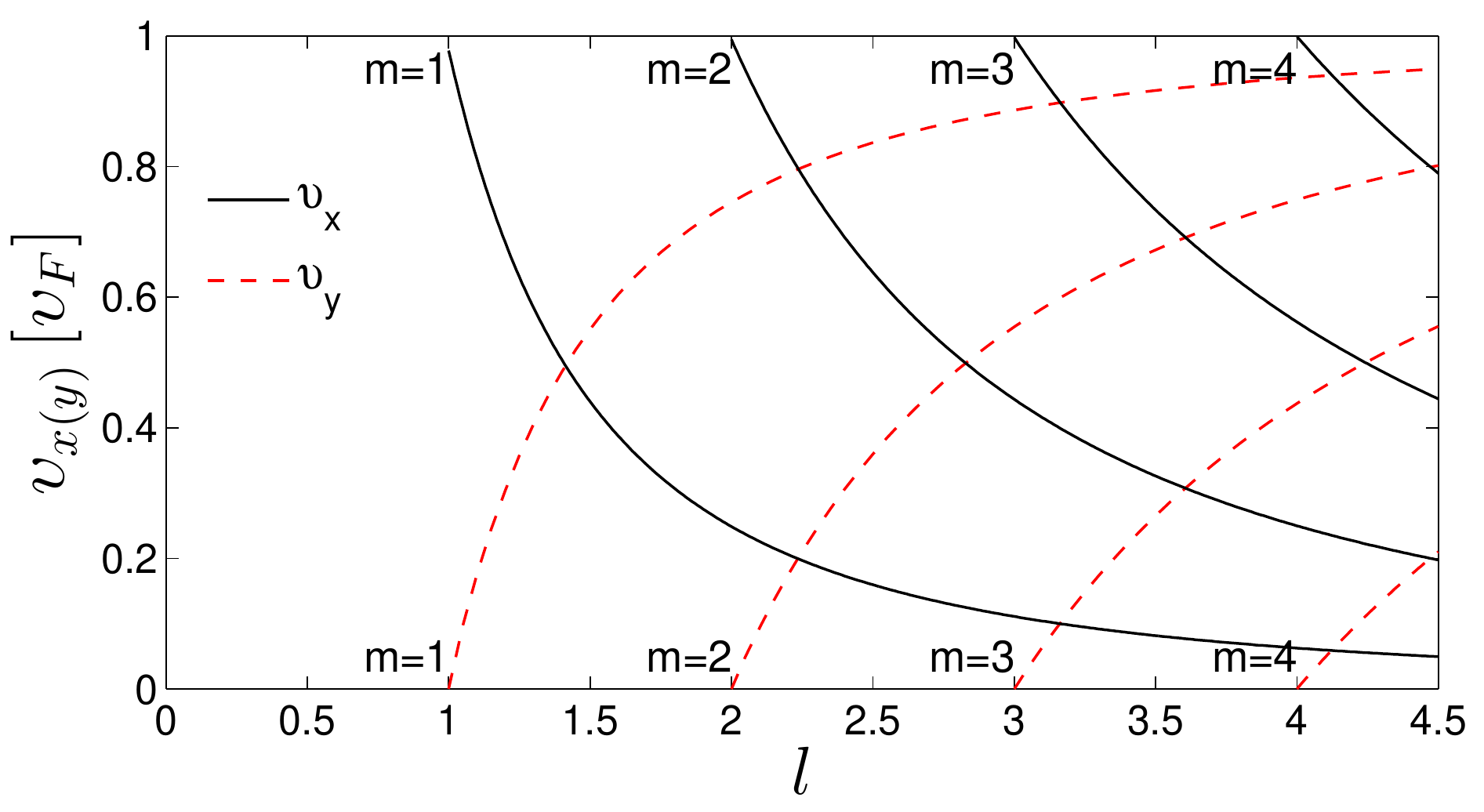}
\caption{(Color online) Group velocities in $\beta_{x}$ and $k_y$ direction vs $l$ at the Dirac point. Parameters are the same as those in Fig. \protect\ref
{fig8}.}\label{fig9}
\end{figure}

In this case, $\frac{{p_{B}}^2+1}{p_{B}}=p_B+\frac{1}{p_B}>2$ (as $p_B\neq1$ and $p_B>0$), $\cos(\o_A-\o_B)\leq1$, and $\cos (\o_A)\cos (\o_B)>0$; therefore the right-hand side of Eq. (\ref{bloch3}) is larger than the integer $1$ unless $-q_Aw_A=q_Bw_B=m\pi$, where $m$ is a positive integer. For the general cases of $-q_Aw_A=q_Bw_B\neq m\pi$, there is no real solution for $\beta_x$. These conditions lead to a gap that is located around the energy satisfying $q_Aw_A+q_Bw_B=0$. If $-q_Aw_A=q_Bw_B=m\pi$, $\beta_x=0$ is the only possible solution. Based on the above analysis and to find zero-energy states, we assume $E=0$ and $\beta_x=0$; thus we have $V_B+\delta<0<V_A$. From $-q_Aw_A=q_Bw_B$ and $\lambda=0.5$, we obtain
\begin{equation}  \label{VB}
V_B=-\sqrt{{V_A}^2+{\delta}^2}.
\end{equation}
From $-q_Aw_A=q_Bw_B=m\pi$, we have
\begin{equation}  \label{kym}
k_{y,\,m}=\pm\sqrt{\frac{{V_A}^2}{(\hbar\upsilon_F)^2}-\frac{{(2m\pi)}^2}{{\Lambda}^2}}.
\end{equation}
Here $m$ denotes the serial number of a Dirac point, which will be explained later. For convenience, we assume $V_A=2l\pi\hbar\upsilon_F/\Lambda=2l\pi\varepsilon_\Lambda$, and Eq. (\ref{kym}) becomes the concise form
\begin{equation}  \label{kylm}
k_{y,\,l,m}=\pm\frac{2\pi}{\Lambda}\sqrt{l^2-m^2}.
\end{equation}
We have $0\leq |k_y|<k_{A}$ for propagating modes; thus $l$ and $m$ satisfy $0<m\leq l$; meanwhile, remember that $m\in\mathbf{Z}$.

Differently, for other cases, i.e., $\lambda\neq0.5$, to find zero-energy states located at $k_{y}=0$, we should consider the conditions $-k_Aw_A=k_Bw_B$ \cite{Wang2010} [$k_{A(B)}$ is the wave vector in the A (B) regions] and $-k_Aw_A=k_Bw_B=m^{\prime}\pi$ ($m^{\prime}\in\mathbf{Z}$), which lead to
\begin{equation}  \label{VBl}
V_{B}^{\prime}=-\sqrt{\frac{{V_{A}^{\prime}}^2{\lambda}^2}{{(\lambda-1)}^2}+{\delta}^2}
\end{equation}
and
\begin{equation}  \label{kylml}
k_{y,\,m^{\prime}}=0,
\end{equation}
with the definition of $V_{A}^{\prime}=m^{\prime}\pi\hbar\upsilon_F/(\lambda\Lambda)$. To find zero-energy states at $k_y\neq0$, we still use $-k_Aw_A=k_Bw_B$ to determine the potentials, and we have $V_{B}^{\prime\prime}=-\sqrt{\frac{{V_{A}^{\prime\prime}}^2{\lambda}^2}{{(\lambda-1)}^2}+{\delta}^2}$ with $V_{A}^{\prime\prime}=l^{\prime\prime}\pi\hbar\upsilon_F/(\lambda\Lambda)$ ($l^{\prime\prime}$ is an arbitrary positive number). Then we assume $-q_Aw_A=m_A\pi$ and $-q_Bw_B=m_B\pi$, where $m_A$ and $m_B$ are both positive integers. From the hypotheses, we have
\begin{eqnarray}  \label{kymab}
k_{y,\,l^{\prime\prime},m_A}=\frac{\pi}{\lambda\Lambda}\sqrt{{l^{\prime\prime}}^{2}-(m_A)^{2}},\notag \\
k_{y,\,l^{\prime\prime},m_B}=\frac{\pi}{(1-\lambda)\Lambda}\sqrt{{l^{\prime\prime}}^2-(m_B)^2}.
\end{eqnarray}
If $k_{y,\,l^{\prime\prime},m_A}=k_{y,\,l^{\prime\prime},m_B}$, there will be a new zero-energy state at $k_{y,\,l^{\prime\prime},m_{A(B)}}$. For $\lambda\neq0.5$, it is difficult to find zero-energy states from Eq. (\ref{bloch2}) using the analytic method; therefore the results stated above are always a convenient approach since the zero-average wave-number gap was found \cite{zk}. In addition, we emphasize that the above solutions for the cases of $\lambda\neq0.5$ are not all the solutions for zero-energy states satisfying $-k_Aw_A=k_Bw_B$, and other unsolved zero-energy states will cluster together with the states we have gotten; hence the obtained solutions are not always Dirac points, which is very different from the case of $\lambda=0.5$.

We continue to discuss the case of $\lambda=0.5$. Figures \ref{fig7}(b) and \ref{fig8}(a)-\ref{fig8}(h) demonstrate the evolution
of the Dirac point versus $l$ for $\lambda=0.5$. Here $l$ can represent $V_A$, as $l$ scales linearly with $V_A$. The value of $l$ and Eq. (\ref{kylm}) determine the number and the positions of the Dirac points. As $l$ increases, Dirac points move away from $k_y=0$, and the Dirac points are generated one by one from $k_y=0$ [see Figs. \ref{fig8}(a)-\ref{fig8}(h)]. If $l<1$, there is no solution for $m$, which means no Dirac points [see Figs. \ref{fig7}(b), \ref{fig8}(a), and \ref{fig8}(b)]. If $l$ is a positive integer, a new Dirac point is generated at $k_y=0$ [see Figs. \ref{fig7}(b), \ref{fig8}(c), \ref{fig8}(e), and \ref{fig8}(g)]. \textcolor{blue}{If} $l$ moves forward from a positive integer, a new pair of Dirac points is generated from $k_y=0$; meanwhile, the Dirac point originally at $k_y=0$ vanishes, which results in an even number of Dirac points [see Figs. \ref{fig7}(b), \ref{fig8}(d), \ref{fig8}(f), and \ref{fig8}(h)].
With fixed $l$, $m$ has possible values $1,2,...,[\,l\,]$, and a larger $m$ denotes a Dirac point or pair of points that is closer to $k_y=0$ or appears later (consider that Dirac point moves away from $k_y=0$ as $l$ grows). For example  [see Fig. \ref{fig8}(g)], with fixed $l=3$, $m=1$ denotes the outermost pair of Dirac points from $k_y=0$, $m=2$ denotes the outer pair of Dirac points from $k_y=0$, and $m=3$ denotes the Dirac point which is exactly located at $k_y=0$. This property is used in Fig. \ref{fig9} to indicate different Dirac points.
Except for creating more Dirac points by increasing $V_A$ and $V_B$ for fixed $\Lambda$, we can achieve a similar result by increasing $\Lambda$ for fixed $V_A$ and $V_B$. In Eq. (\ref{kym}), if we increase $\Lambda$, we will get more reasonable solutions of $m$, i.e., more Dirac points appearing. Under this condition, $m$ still must be a positive integer.

\begin{figure*}[b t p]
\centering
\includegraphics[width=17cm]{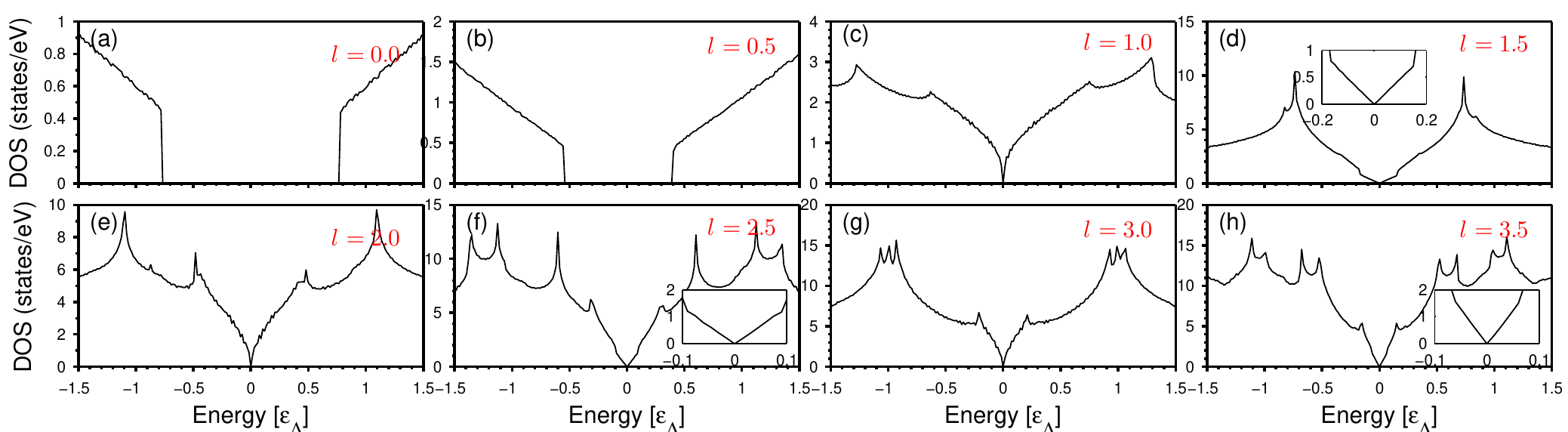}
\caption{(Color online) Density of states for different $l$. The other parameters are the same as those in Fig. \protect\ref
{fig8}.}\label{fig10}
\end{figure*}

We plot the DOS in Figs. \ref{fig10}(a)-\ref{fig10}(h), which correspond to Figs. \ref{fig8}(a)-\ref{fig8}(h), respectively. When zero-energy modes do not exist, as in Figs. \ref{fig8}(a) and \ref{fig8}(b), the DOS is zero around zero energy. As for cases with $l\geq1$, zero-energy modes emerge. Nevertheless, the relations between the DOS and the energy are different for different $l$ near zero energy. If $l$ is a positive integer, as in Figs. \ref{fig10}(c), \ref{fig10}(e), and \ref{fig10}(g), a more or less curvilinear rise in the DOS versus the energy can be observed. For the other cases, i.e., Figs. \ref{fig10}(d), \ref{fig10}(f), and \ref{fig10}(h), the DOS increases linearly. The linear relation between the DOS and the energy is formed because of the extra linearlike Dirac cones around zero energy, although these extra linearlike Dirac cones may be anisotropic. At each Dirac point, the group velocity becomes zero in a few directions; therefore there is no Van Hove singularity at zero energy.

\begin{figure}[h p t b ]
\centering
\includegraphics[width=7.5cm]{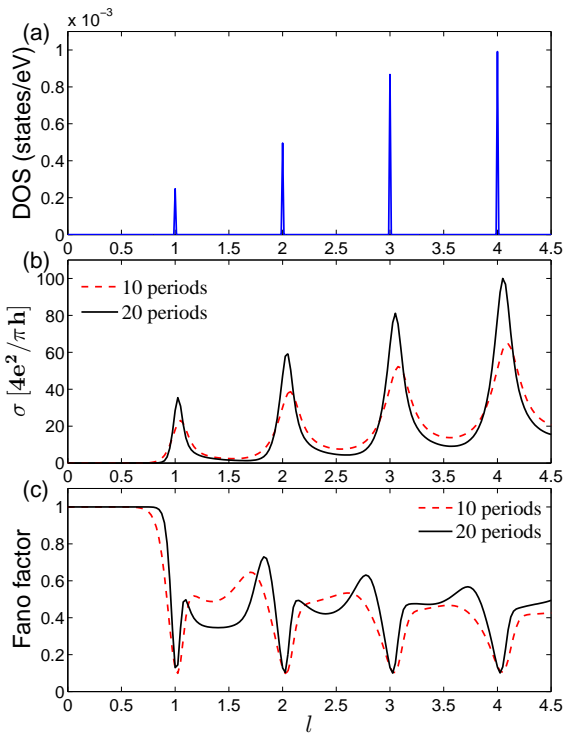}
\caption{(Color online) (a) Density of states, (b) conductivity, and (c) Fano factor at zero energy vs $l$. For (b) and (c), $L_y/L_x\gg 1$. The other parameters are identical to those in Fig. \protect\ref{fig8}.}
\label{fig11}
\end{figure}

\begin{figure}[ !h p b]
\centering
\includegraphics[width=7.5cm]{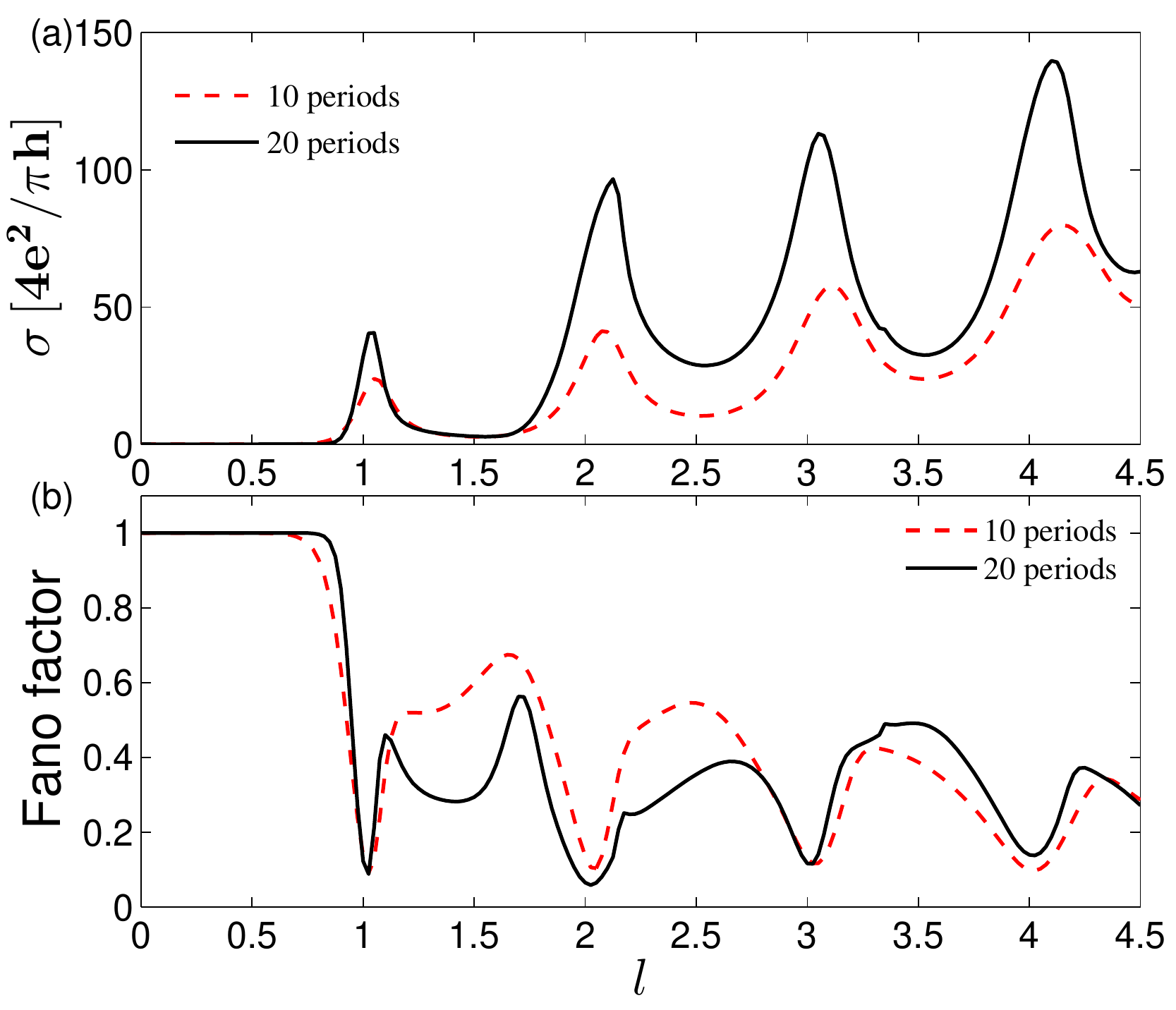}
\caption{(Color online) Effect of the structural disorder on (a) the conductivity and (b) the Fano factor with $w_A=(20+R_2)$ nm and $w_B=(20+R_2)$ nm, where $R_2$ is a random number between $-1$ and $+1$ nm. The other parameters are the same as those in Fig. \ref{fig11}.}
\label{fig12}
\end{figure}

Inspired by the work on the group velocity at the Dirac point of pristine graphene \cite{Park2009,Barbier2010,Park2008}, now we discuss that in this structure, which helps us to understand the resonance in Fig. \ref{fig11}. Group velocities in the $\beta_{x}$ and $k_y$ directions can be given by
\begin{equation}  \label{velocity}
\upsilon_{x}=\frac{\partial E}{\partial p_{x}}, \quad\upsilon_{y}=\frac{\partial E}{\partial p_{y}},
\end{equation}
where $p_{x}=\hbar \beta_{x}$ and $p_{y}=\hbar k_{y}$ are the momenta in the $\beta_{x}$ and $k_y$ directions in terms of the Bloch wave. It is not convenient to find the analytical solution of Eq. (\ref{velocity}) from Eq. (\ref{bloch2}); therefore we present the result in Fig. \ref{fig9} numerically. As stated previously, the Dirac point at $k_y=0$ exists only when $l$ is a positive integer. Figure \ref{fig9} shows that every time it appears, $\upsilon_x\simeq\upsilon_F$ and $\upsilon_y=0$. When the Dirac point at $k_y=0$ vanishes, as $l$ increases, the corresponding $\upsilon_x$ of the new pair of Dirac points gradually reduces to zero from approximately $\upsilon_F$, and $\upsilon_y$ gradually approaches $\upsilon_F$ from zero. The cases with $m=1,2,3$, and $4$ are shown in Fig. \ref{fig9}; as can be seen, they all have similar characters.

The DOS, the conductivity, and the Fano factor at zero energy are shown in Fig. \ref{fig11}. Brey and Fertig have found conductance resonance in the uniform gapless graphene with the periodic cosine-type potential \cite{Brey2009}; here we observe a similar property as well. From Fig. \ref{fig11}, it can be seen that there are peaks in the conductivity and valleys in the Fano factor when $l$ are positive integers, and small values (about 0.1) of the Fano factor suggest the resonance at zero energy. The resonance may be caused by the appearance of the Dirac point at $k_y=0$ and its zero group velocity along the $k_y$ direction, as the new Dirac point and the character of $\upsilon_y=0$ lead to the strong enhancement of the DOS near zero energy [see Fig. \ref{fig11}(a)]. It can also be seen intuitively from Figs. \ref{fig8}(c), \ref{fig8}(e), and \ref{fig8}(g) that the band at ($k_y=0$, $E=0$) becomes flat in $k_y$ direction, which results in the enhancement of the DOS. In addition, compared with Fig. \ref{fig5}(b), the conductivity versus $l$ in Fig. \ref{fig11}(b) has the same rising tendency if we ignore the resonance, which indicates that the potential (represented by $l$) enhances the conductivity by generating more Dirac points at zero energy.

Then we consider the effect of the structural disorder on the conductivity and the Fano factor at zero energy. The percentage of the largest width derivation $R_2$ against $w_A/w_B$ is up to $5\%$, and we show the conductivity and the Fano factor under such structural disorder in Fig. \ref{fig12}. Figure \ref{fig12} explicitly shows that peaks of the conductivity and valleys of the Fano factor still exist, although the shift of these peaks becomes more distinct.

Finally, in experiments, we suggest that our results could be realized on a heterosubstrate such as SiO$_2$/BN \cite{substrate}, and the conditions are that one part of the inhomogeneous substrate breaks the
sublattice symmetry, which leads to a symmetry gap around the Fermi level, and another part does not. In addition, other devices which have a similar principle may be equally valid. For example, graphene with a band gap induced by patterned hydrogen adsorption \cite{Balog2010} can take the place of graphene on the BN substrate.

\section{Conclusions}\label{part4}
In summary, we investigated the band gap around the Fermi level and zero-energy modes of the electronic band structures of $1$D graphene-based superlattices placed on the heterosubstrate with periodic square potentials.

It is found that the band gap's width can be almost linearly tuned by the proportion of inhomogeneous substrate if equal potentials are applied on the GSLs, and the relation is robust against the effect of the structural disorder. The relation between the band gap and the proportion of an inhomogeneous substrate is exactly like comproportionation \cite{comproportionation} in chemistry, which will benefit the design of graphene-based electronic devices. Moreover, the scaling law of the band gap is explained by the wave function and the DOS.

For zero-energy modes, the typical case of equal widths of an inhomogeneous substrate was discussed in detail. Although the sublattice symmetry was broken for gapped graphene, we showed that Dirac points emerge at zero energy if asymmetric potentials are applied on the GSLs, but the Dirac point at $\mathbf{k}=\mathbf{0}$ [\,as $(\beta_x,k_y)=(0,0)$\,] exists only for specific potentials. Once the Dirac point at $\mathbf{k}=\mathbf{0}$ appears, the resonance occurs with the conductivity having a peak and the shot noise tending to be very low, which indicates that the transport becomes ballistic. Furthermore, we found that the resonance occurs for the strong enhancement of the DOS around zero energy, which is caused by the appearance of a Dirac point at $\mathbf{k}=\mathbf{0}$ and its zero group velocity in the $k_y$ direction. General cases with unequal widths of the inhomogeneous substrate were also discussed, and part of the zero-energy states was described analytically. Our prediction may be realized on a heterosubstrate such as SiO$_{2}$/BN, and other devices which obey a similar principle should be equally possible.

\section{Acknowledgments}\label{part5}
T.M. thanks CAEP for partial financial support. This work is supported by NSFC (Grants No. 11274275 and No. 11374034) and the National Basic Research Program of China (Grants No. 2011CBA00108 and No. 2012CB921602). We also acknowledge the support from the Fundamental Research Funds for the Center Universities under Grants No. 2015FZA3002, No. 2016FZA3004 (L.-G.W.), and No. 2014KJJCB26 (T.M.), the HSCC of Beijing Normal University, and the Special Program for Applied Research on Super Computation of the NSFC-Guangdong Joint Fund (the second phase).

\end{document}